\documentstyle[twoside,fleqn,espcrc2,epsfig]{article}
\input {epsf.sty}
\input {psfig.sty}

\newcommand{\AmS}{{\protect\the\textfont2
A\kern-.1667em\lower.5ex\hbox{M}\kern-.125emS}}

\title{A Note on the Action in $d > 4$ Dynamical Triangulations}

\author{Alun George
\address{Department of Physics, University of Wales Swansea, 
Singleton Park, Swansea SA2 8PP, UK. }}
        
\begin{document}

\begin{abstract}
For dynamical triangulations in dimensions $d\leq4$ the most general action has two couplings. We note that the most general action for $d=5$ has three couplings. We explore this larger coupling space using Monte Carlo simulations. Initial results indicate evidence for non-trivial phase structure.
\end{abstract}

\maketitle

\section{INTRODUCION}

Dynamical triangulations (DT) have been widely studied in the past decade as a non-perturbative model of quantum gravity. Naturally, most of the work done has been in dimensions $d = 2$, $3$ and $4$. The action used to describe pure gravity in Euclidean $d$-space is the Einstein-Hilbert action

\begin{equation}
S[g] = \frac{1}{16 \pi G} \int d^{d}x \sqrt{g}(2\Lambda - R),
\end{equation}
where $\Lambda$ is the cosmological constant, $R$ is the scalar curvature, $G$ is Newton's constant and $g$ is the spacetime metric. The discretized version of the continuum action is well known and can be written as 

\begin{equation}
S[\tau] = \kappa_d N_d - \kappa_{d-2} N_{d-2},
\end{equation}
where $N_i$ is the total number of $i$-(sub)simplices in the simplicial manifold $\tau$, $\kappa_d$ and $\kappa_{d-2}$ represent the cosmological constant and Newton's constant respectively. The partition function (\ref{eq:part}) can then be defined as the sum over all possible triangulations $\tau$ of a manifold of given topology, each weighted by its Boltzmann factor. The topology is normally fixed to a $d$-sphere for simplicity.

\begin{equation}
Z(\kappa_{d}, \kappa_{d-2)} = \sum_{\tau : S^{d}} exp(\kappa_{d-2} N_{d-2} - \kappa_d N_d ) \label{eq:part}
\end{equation}

\section{EQUIVALENT ACTION}

It can be convenient to express $S$ in a different but equivalent form $S'$ \cite{1}.

\begin{equation}
S' = \kappa_{d}' N_d - \kappa_{0} N_{0},
\end{equation}
where $\kappa_{d}' \neq \kappa_d$ and $N_0$ is the number of nodes in the triangulation. This is sometimes done since this action can be easier to use in a Monte Carlo algorithm.

The actions $S$ and $S'$ can be shown to be equivalent in dimensions 2, 3 and 4 using the Dehn-Sommerville (\ref{eq:dehn}) and Euler (\ref{eq:euler}) relations. These are simple linear relations between the total number of (sub)simplices ($N_i$) in a given simplicial manifold. In $d$-dimensions these relations are written as

\begin{equation}
N_i = \sum_{j=i}^{d} (-1)^{d-j} \left(\begin{array}{c} j+1 \\ i+1 \end{array}\right) N_j \label{eq:dehn}
\end{equation}
and

\begin{equation}
\chi = \sum_{i=0}^{d} (-1)^{i} N_i, \label{eq:euler}   
\end{equation}
where $\chi$ is the Euler characteristic ($\chi=0$ for $S^{5}$ topology). 
 
The equivalence of $S$ and $S'$ boils down to the question of whether one can express $N_0$ as a function of $N_d$ and $N_{d-2}$. This is indeed possible in $d$ = 2, 3 and $4$. In 5$d$ equations (\ref{eq:dehn}) and (\ref{eq:euler}) reduce to the following set of three independent relations.

\begin{displaymath}
N_0 - N_1 + N_3 - 3N_5 = 0
\end{displaymath}

\begin{equation}
N_2 - 2N_3 + 5N_5 = 0 \label{eq:indep}
\end{equation}

\begin{displaymath}
N_4 - 3N_5 = 0
\end{displaymath}
Clearly $N_0$ cannot be expressed in terms of $N_5$ and $N_3$ alone. Therefore $S'$ is {\em not} equivalent to $S$ in $d = 5$. Similar results are found in $d > 5$. This result forces us to use the action 

\begin{equation}
S = \kappa_5 N_5 - \kappa_3 N_3 \label{eq:k3action}
\end{equation}
in five dimensions rather than

\begin{equation}
S' = \kappa_{5}' N_5 - \kappa_0 N_0 \label{eq:k0action}
\end{equation}
if we intend studying what we believe to be dynamically triangulated gravity, using the Einstein-Hilbert action on the lattice. 

\begin{figure}
\centerline{
\setlength\epsfxsize{230pt}
\epsfbox{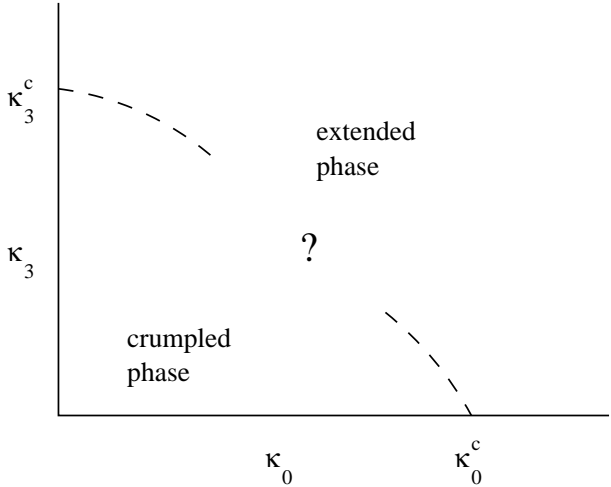}
}
\vspace{-8mm}
\caption{$\kappa_3$-$\kappa_0$ surface phase diagram at $\kappa_{5}^{c}$ in the three dimensional coupling constant space.}
\label{K0K3}
\vspace{-5mm}
\end{figure}

From equations (\ref{eq:indep}) we can see that the most general action linear in $N_{i}$ will contain three terms since we have six variables related by three independent equations. The coupling constant space is therefore three dimensional (see fig.\ \ref{K0K3}). The special case of action (\ref{eq:k3action}) would represent a surface in this three dimensional space. The most general action $S_{gen}$ in 5$d$ could have the form 

\begin{displaymath}
S_{gen} = \kappa_5 N_5 - \kappa_3 N_3 - \kappa_0 N_0. \label{eq:S-gen}
\end{displaymath}
Such a situation does not arise in $d\leq4$ because the most general actions are two dimensional. So five is the lowest dimension in which this property exists. 

\section{$d > 4$}

The aim of this work is to map out the three dimensional coupling constant space of 5$d$ DT using $S_{gen}$ in the hope that it will reveal some extra phase structure. Monte Carlo simulations are first run using $S_{gen}$ with $\kappa_{0}=0$ (tetrahedral action) and $\kappa_{3}=0$ (nodal action). The observables measured include the average curvature $\langle R \rangle = \langle N_{d-2}/N_{d} \rangle$, its susceptibility $\langle R_{sus} \rangle$ and the average geodesic length $\langle d \rangle$. A phase transition would result in a sharp rise in $\langle R_{sus} \rangle$ since it is a second derivative of the free energy. 

If both limits produce identical phase transitions and are in the same universality class, then this forces the question: what is so special about the action that we derived from the continuum Einstein-Hilbert action? It would also be interesting if both limits have distinct phase transitions since this would tell us that there could well be something special about our derived action. 

\section{RESULTS}

\begin{figure}
\centerline{
\setlength\epsfxsize{230pt}
\epsfbox{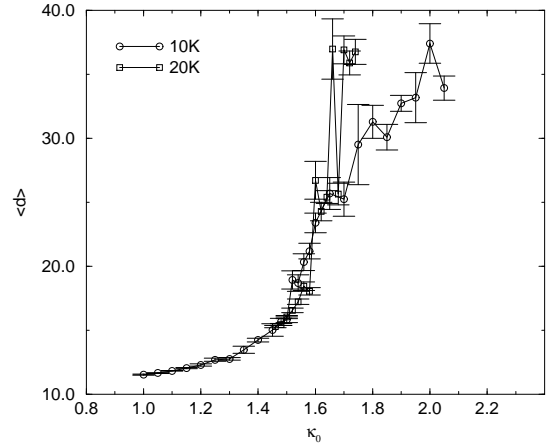}
}
\vspace{-12mm}
\caption{Nodal action: crumpled (small $\kappa_0$) and extended phases (large $\kappa_0$).}
\label{K0-d}
\vspace{-6mm}
\end{figure}

\begin{figure}
\centerline{
\setlength\epsfxsize{230pt}
\epsfbox{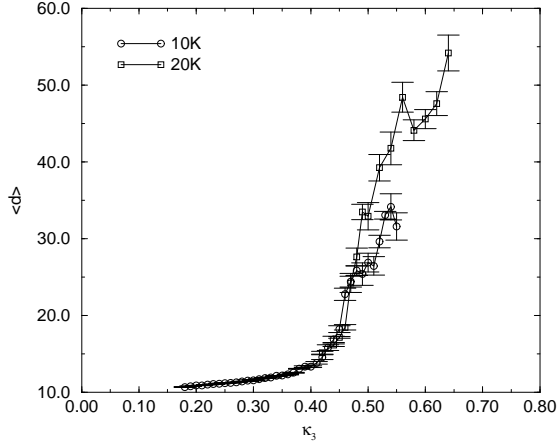}
}
\vspace{-12mm}
\caption{Tetrahedral action: crumpled (small $\kappa_3$) and extended phases (large $\kappa_3$).}
\label{K3-d}
\vspace{-5mm}
\end{figure}

So far, we have identified phase transitions in both models. These are evident from the sharp rise in the geodesic lengths (see figs. \ref{K0-d} and \ref{K3-d}), possibly indicating a branched polymer phase as found in 4$d$, (but this would require further study to be confirmed); and small peaks in the curvature susceptibility $\langle R_{sus} \rangle$.

\begin{figure}
\vspace{-8mm}
\centerline{
\setlength\epsfxsize{230pt}
\epsfbox{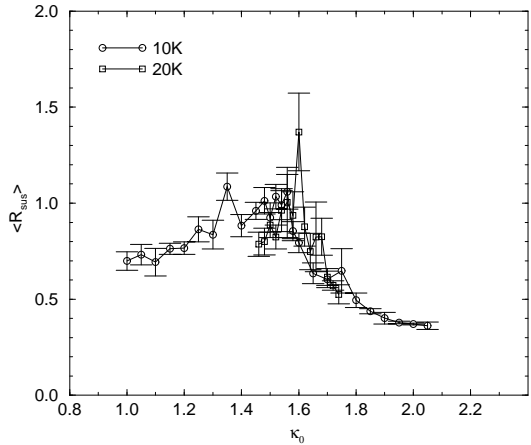}
}
\vspace{-15mm}
\caption{Nodal action: Peak in the curvature susceptibility.}
\label{K0-Rsus}
\vspace{-10mm}
\end{figure}

\begin{figure}
\centerline{
\setlength\epsfxsize{230pt}
\epsfbox{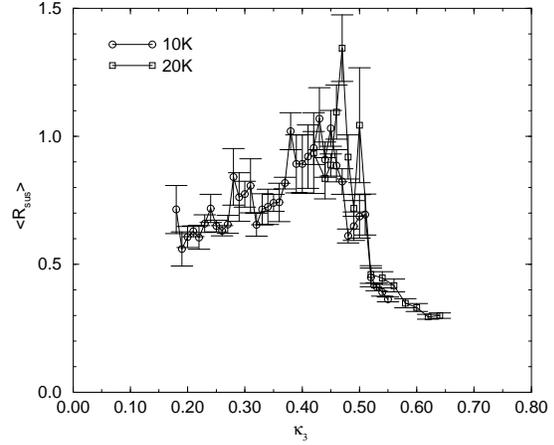}
}
\vspace{-12mm}
\caption{Tetrahedral action: Peak in the curvature susceptibility.}
\label{K3-Rsus}
\vspace{-5mm}
\end{figure}

The size of the peak itself is relatively small, due to finite size effects of simulating lattice volumes of 10K and 20K, which correspond to a lattices of $\sim 6^5$ and $\sim 7^5$ respectively. The {\em approximate} location of the phase transitions were identified by running Monte Carlo simulations of $10^5$ sweeps. Longer runs of $5\times10^5$ sweeps were then done near the phase transitions in order to reduce errors. The phase transition for the nodal coupling was near $1.6 \pm 0.1$ (see fig.\ \ref{K0-Rsus}) and at $0.45 \pm 0.05$ for the tetrahedral coupling (see fig.\ \ref{K3-Rsus}).

\section{FURTHER STUDY}

Once the order of the phase transitions and critical exponents are known, the next stage would be to investigate the region `between' these two limits (see fig.\ \ref{K0K3}), ie. where $\kappa_3$ and $\kappa_0$ are non-zero in $S_{gen}$. This would give us a complete picture of the phase space. It will also be interesting to measure the distribution of singular vertices across the 3$d$ coupling constant space in order to compare results with \cite{1}. 

\vspace{5mm}

{\bf Acknowledgements}
We thank Simon Catterall for a copy of his DT code and Simon Hands for useful discussions.

\end{document}